\documentstyle[12pt,psfig,epsfig]{article}
\newcommand{\select}[2]{#1}

\begin{document}

\begin{flushright}
hep-th/0307023
\end{flushright}

\def\be{\begin{equation}}
\def\ee{\end{equation}}

\vspace{5mm}

\begin{center}
{\Large \bf Brane World of Warp Geometry:\\[1mm]
An Introductory
Review\footnote{Review paper}}\\[13mm]
Yoonbai Kim,~~
Chong Oh Lee,~~
Ilbong Lee
\\[1mm] \em
BK21 Physics Research Division and Institute of Basic Science,
Sungkyunkwan University,\\
Suwon 440-746, Korea\\
{\tt yoonbai$@$skku.ac.kr, cohlee$@$newton.skku.ac.kr,
ilbong$@$newton.skku.ac.kr}\\
[7mm] \em
JungJai Lee
\\[1mm] \em
Department of Physics, Daejin University, GyeongGi, Pocheon 487-711, Korea\\
Department of Physics, North Carolina State University, Raleigh, NC 27695-8202, USA\\
{\tt jjlee$@$daejin.ac.kr}
\end{center}

\vspace{5mm}

\begin{abstract}
Basic idea of Randall-Sundrum brane world model I and II is reviewed with
detailed calculation. After introducing the brane world metric with
exponential warp factor, metrics of Randall-Sundrum models are
constructed. We explain how Randall-Sundrum model I with two branes
makes the gauge hierarchy problem much milder, and derive
Newtonian gravity in Randall-Sundrum model II with a single brane
by considering small fluctuations.
\end{abstract}


\newpage

\setcounter{equation}{0}
\section{Introduction}
During the last few years the brane world scenario inspired by
developments
in string theory has
attracted much attention in particle physics, cosmology, and astrophysics.
Basic structure of the brane world scenario is understood by two
representative models. One is Arkani-Hamed-Dimopoulos-Dvali
model~\cite{ADD} and the other is Randall-Sundrum (RS) brane world
models I and II~\cite{RS1,RS2}.
The main purpose of this pedagogical review is to introduce the original
form of RS models as precise as possible despite of numerous
results~\cite{Rev} in diverse research
directions~\cite{KK,GW,GRS,BDL,Od,ABN,CCV,CK,APR,ADKS,KIS}.

The motivation of RS model I is to propose a resolution of the
gauge hierarchy problem, a long standing puzzle in particle
phenomenology, from the viewpoint based on the geometry of our
spacetime structure instead of symmetry principle like
supersymmetry. Here let us briefly explain what the gauge
hierarchy problem is. According to the standard model employing
the idea of gauge symmetry and its spontaneous breaking, the mass
scale of electroweak symmetry breaking is $M_{\rm EW}\sim 10^{3}$
GeV which means each gauge particle has mass of order
$10^{-24}$kg but that of gravity is the Planck scale $M_{\rm
Planck}\sim 10^{19}$ GeV. For the units and conversion factors,
refer to Tables 1 and 2
in Appendix A. This huge gap between the electroweak scale and
the Planck scale, $M_{\rm EW}/M_{\rm Planck}\sim10^{-16}$, needs
a fine tuning up to 16 digits.

Let us understand
the meaning of the fine tuning by using a toy example. Suppose we
observe a particle of mass $m_{{\rm experiment}}\approx 1,100$ GeV through
experiments. However, quantum field theory computation usually predicts
enormous quantum correction like $\Delta m_{{\rm quantum~correction}}
\sim 10^{19}$ GeV irrespective of the bare mass parameter $m_{{\rm bare}}$,
which coincides with the ultraviolet cutoff in order $M_{\rm Planck}$.
Since we can regard this bare mass parameter as classical
mass of a particle in the classical Lagrangian,
a natural bare mass parameter should be about
$m_{{\rm bare}}\sim
m_{{\rm experiment}}\approx 1,100$ GeV in the environment of the
electroweak scale. On the other hand, a simple but unavoidable algebra requires
that $m_{{\rm bare}}$ is not $m_{{\rm experiment}}\approx 1,100$ GeV but
$m_{{\rm bare}}\approx m_{{\rm experiment}}-\Delta m_{{\rm quantum~correction}}
\sim 1.1\times 10^{3}-10^{19}$ GeV.
A fine tuning of $m_{{\rm bare}}$ up to 16 digits
like $m_{{\rm bare}}=-9.999999999999989 \times 10^{18}$ GeV is a nonsense
in any rational science. It means that the standard model at present form seems
imperfect and this gauge hierarchy problem hinders unifying the standard model
in electroweak scale and the gravity in Planck scale.
Thus we need an additional physical principle to protect physical
results from the above nonsensical fine tuning.
We will introduce the RS brane world model I~\cite{RS1}
in subsection 3.2, and explain
how the warp factor in the RS I makes the gauge hierarchy problem much
milder without introducing other ingredients like supersymmetry
in subsection 4.2.

The RS models are constructed in the scheme of general relativity so that
the gravity induced on the 3-brane(our universe) 
should satisfy the observational and
experimental bounds. The first step is the reproduction of Newtonian
gravity
on the 3-brane in the weak gravity limit with no doubt. Though it seems
nontrivial due to negative cosmological constant in the bulk, the induced
gravity on the 3-brane in RS II is exactly the Newtonian gravity
from the zero mode of small gravitational fluctuations, and the small
corrections are given by continuous tower of
higher Kaluza-Klein(KK) modes~\cite{RS2}.

The rest of the paper is organized as follows. In section 2, we
introduce a few basic ingredients in general relativity for
subsequent sections, including the metric, Einstein-Hilbert
action, cosmological constant, Einstein equations, and
Kretschmann invariant. Section 3 is composed of 3 subsections. In
subsection 3.1, we compute some properties of 5-dimensional pure
anti-de Sitter spacetime. In subsections 3.3 and 3.2, we give a
detailed description of the geometry of RS model I with two
3-branes and RS model II with the single 3-brane, respectively.
In subsection 4.1, we consider small gravitational fluctuations
on the 3-brane in RS model II and show that their zero mode
depicts the Newtonian gravity. In subsection 4.2, we show how to
treat the gauge hierarchy problem in the scheme of RS I by using
the warp factor. We firstly derive 4-dimensional gravity on our
3-brane, and then demonstrate the emergence of the electroweak
scale masses for Higgs, gauge boson, and fermion. We conclude in
section 5 with a summary and an introduction of viable research
directions of RS models I and II.

\setcounter{equation}{0}
\section{Setup}

In order to study and construct various brane world scenarios with
warp factor, as a basic language, the general relativity is good. This
seems
indispensable since the description of the early universe has been made
by the cosmological solutions of Einstein equations. In this section we
introduce
a minimal setup and basic notions for the brane world scenarios.
Definitions and notations we use are summarized in Appendix A,
and the detailed derivation of various equations and quantities are given
in
Appendix B.

\begin{figure}[ht]
\centerline{\scalebox{1}{\includegraphics{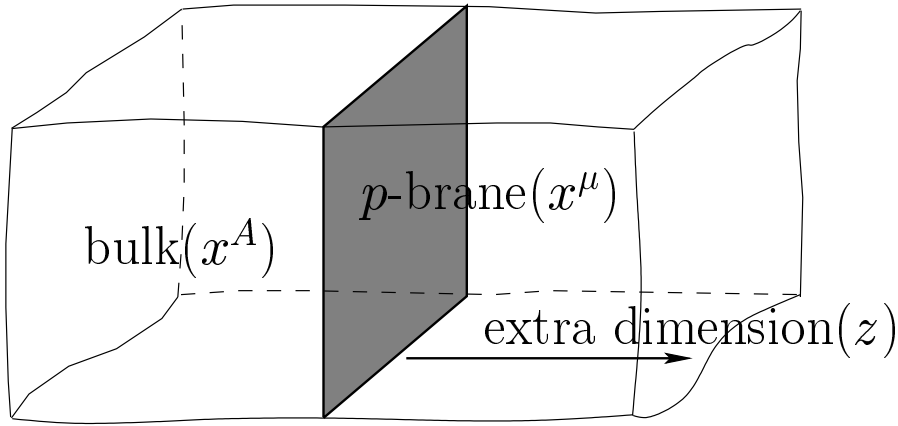}}}
\caption{A schematic shape of brane world model: Spatial section of our
universe at time $t$ is a {\it brane} (shaded region) expressed by
coordinates $x^{i}$ and that of higher dimensional {\it bulk} embedding our
universe (transparent box) is depicted by $x^{A}$.
One coordinate $z$ transverse to the brane is {\it extra dimension.}}
\label{rsrfig0}
\end{figure}

In $D$-dimensional curved spacetime composed of a time $t$, a $p$-brane
$x^{i}$, and an extra-dimension $z$, the geometry of the curved spacetime
is
described by the metric
\begin{eqnarray}
ds^{2}&=&g_{AB}dx^{A}dx^{B} \\
&=&g_{\mu\nu}dx^{\mu}dx^{\nu}+2g_{\mu D}dx^{\mu}dz+g_{DD}dz^2 \\
&=&g_{00}dt^{2}+2g_{0i}dtdx^{i}+g_{ij}dx^{i}dx^{j}+2g_{\mu
D}dx^{\mu}dz+g_{DD}dz^2.
\label{met5}
\end{eqnarray}
From here on, the capital Roman indices ($A, B, \cdots =0,1,\cdots, p,
p+1$)
denote $D$-dimensional bulk spacetime indices $(D=p+2)$,
the Greek indices ($\mu, \nu,
\cdots =0,1,\cdots, p$) the spacetime indices of the worldbrane, and small
Roman
indices ($a, b, \cdots i,j,k, \cdots =1,2,\cdots, p$) the coordinates of
the brane. Therefore, we call the space described by the coordinates
transverse
to the $p$-brane is called by {\it extra dimensions}.
Obviously, the main concern is our world of $p=3$ since our present
spacetime is (1+3)-dimensional and the extra dimension is one denoted by
$z$-coordinate as the simplest case.
A schematic shape of the brane world is shown in Fig.~\ref{rsrfig0}.

The action  of our interest is
\begin{equation}\label{action}
S = \int d^{D}x \sqrt {g_{D}} \left[-\frac {M^{p}_{*}}{16 \pi}
(R+2\Lambda)\right] + S_{\rm matter},
\end{equation}
where $M_{*}$ is the fundamental scale of the theory, $\Lambda$ a
cosmological constant, and $S_{{\rm matter}}$ stands for any matter
of our interest.
We read Einstein equations in the ($p$+2)-dimensional
bulk from the action (\ref{action})
\begin{equation}\label{ein}
G_{AB} \equiv R_{AB} - \frac{1}{2} g_{AB}R = -\frac {8\pi}{M^{p}_{*}}
T_{AB} + g_{AB} \Lambda ,
\end{equation}
where energy-momentum tensor $T_{AB}$ in Eq.~(\ref{ein}) is defined by
\begin{eqnarray}\label{emt}
T_{AB}\equiv  \frac{2} {\sqrt{g_D}} \,\,  \frac{\delta S_{\rm
matter}}{\delta g^{AB}}.
\end{eqnarray}

In this pedagogical
review, we take into account the matters restricted to the brane, which
coincide with those of original idea~\cite{RS1,RS2}.
When the matters are confined on a specific $p$-brane, then the metric
fluctuation to the $z$-direction so-called radion direction vanishes,
i.e., $\delta g^{zz}=0$, and
thereby the energy-momentum tensor from the sources confined on the brane
becomes
\begin{equation}
T_{AB}^{\rm brane}\equiv \left. \frac{2} {\sqrt{g_D}} \,\,  \frac{\delta S_{\rm
matter}^{\rm brane}}{\delta g^{AB}} \right|_{\delta g^{zz}=0}
.\label{emtb}
\end{equation}

When we particularly consider the metric ansatz without the
 cross terms among time variable $t$, spatial coordinates of
the brane $x^{i}$, $(i=1, 2, 3)$, and coordinate of the extra-dimension $z$,
a 5-dimensional metric is expressed as follows, which includes a flat
$p=3$
brane and is convenient for the description of the brane world
\begin{equation}\label{5dmet}
ds^2 = e^{2A(t,z)}[dt^2 - D^{2}(t,z)dx^{i2}] - C^{2}(t,z)dz^{2},
\end{equation}
where $A(t,z)$, $D(t,z)$, and $C(t,z)$ are three real metric functions of $t$
and $z$~\cite{KK}. Actually, vanishing off-diagonal metric
components in front of $dx^{\mu}dz$ is consistent with the reflection
symmetry
of the $z$-coordinate for orbifold compactification, i.e., $z\rightarrow
-z$. Similar symmetry argument, e.g., time-reversal ($t \rightarrow -t$)
or parity ($x^{i}\rightarrow -x^{i}$ for $p$ = 3), is also applied
to the $p$-brane, which results
in vanishing $dtdx^{i}$ component. If the geometry of our interest is
static, Eq.~(\ref{5dmet}) becomes
\begin{equation}\label{u1s}
ds^2 = e^{2A(z)}[dt^2 - D^{2}(z)dx^{i2}] - C^{2}(z)dz^{2}.
\end{equation}
Introducing a new coordinate $Z$ such as $dZ=C(z)dz$, we rewrite
the metric (\ref{u1s}) as
\begin{equation}\label{u2s}
ds^2 = e^{2A(z(Z))}[dt^2 - D^{2}(z(Z))dx^{i2}] - dZ^{2}.
\end{equation}
Eq.~(\ref{u2s}) has two independent metric functions.
If we force Poincar\'{e} symmetry with the unit light speed
for the spacetime of the $p$-brane,
then the boost symmetry asks $D^2(Z(z))=1$ so that
we finally arrive at
\begin{equation}\label{stamet}
ds^2 = e^{2A(Z)}(dt^2 - dx^{i2}) - dZ^{2} =
e^{2A(Z)}\eta_{\mu\nu}dx^{\mu}dx^{\nu}-dZ^2.
\end{equation}

On the other hand, when the cosmology is our interest, we have to consider
the homogeneous and isotropic $p$-brane. The simplest model is depicted
by
the metric which involves time-dependence only in front of the 3-brane
coordinates, i.e., $D(t,Z)=e^{b(t)}$ :
\begin{equation}\label{cosmet}
ds^2 = e^{2A(Z)}[dt^2 - e^{2b(t)}dx^{i\,2}]- dZ^{2},
\end{equation}
which leads to Eq.~(\ref{stamet}) in static limit.
If a constant curvature consistent with the homogeneity and isotropy is
included, we have
\begin{equation}\label{come}
ds^2 = e^{2A(Z)}\left[dt^2 - e^{2b(t)}\left(\frac{dr^{2}}{1-Kr^{2}}
+r^{2}d\theta^{2}+r^{2}\sin^{2}\theta d\phi^{2}\right)\right]- dZ^{2},
\end{equation}
where $K=1$ corresponds to three sphere of unit radius, $K=0$
3-dimensional flat space, and $K=-1$ three hyperbolic space.

For this metric (\ref{stamet}), the Einstein equations (\ref{ein}) are
given by the following simple equations
\begin{eqnarray}
A^{'2}&=&\frac{2}{p(p+1)}\left(\frac{8\pi}{M^{p}_{\ast}}T^{Z}_{\;Z}
-\Lambda\right),
\label{sezi}\\
A''&=&-\frac{1}{p}\frac{8\pi}{M_{*}^{p}}\left(T_{\;Z}^{Z} - T_{\;t}^{t}
\right),\qquad \left(T^{t}_{\;t}=T^{i}_{\;i}\right),
\label{sezt}
\end{eqnarray}
where the prime in ${A}^{'}$ denotes the differentiation by
$Z$-coordinate. In order to identify the physical singularity, we
look into sum of square of all components of the Riemann
curvature tensor so-called the Kretschmann scalar invariant from
the metric (\ref{cosmet})
\begin{equation}\label{KrA}
R^{ABCD}R_{ABCD}=2(p+1)\left[pA^{'4}+2\left(A''+A'^{2}\right)^2\right].
\end{equation}
Derivation of the above equations and quantities are given in Appendix B.

A warp coordinate system (\ref{cosmet}) is unusual for the description of
anti-de Sitter spacetime so that we introduce familiar logarithmic coordinate
such as $dZ\sim \pm dy/\sqrt{B(y)}$ with $B(y)\sim e^{2A(Z)}$. Then the
metric (\ref{cosmet}) is rewritten in other coordinates
\begin{equation}\label{metric1}
ds^{2} = B(y) (dt^{2}-dx^{i2})-\frac{dy^2}{B(y)},
\end{equation}
and corresponding Einstein equations are
\begin{eqnarray}
B^{'^2} &=& \frac{8B}{p(p+1)} \left( \frac{8\pi}{M_{*}^{p}} T^{y}_{y} -\Lambda
\right ),
\label{etzi}\\
B^{''}&=& -\frac{p+2}{p(p+1)}\frac{32\pi}{pM_{*}^{p}}
\left( T^{y}_{y} - T^{t}_{t} \right)
+\frac{4\Lambda}{p(p+1)},\qquad \left(T^{t}_{t}= T^{i}_{i}\right),
\label{ettz}
\end{eqnarray}
where the prime ${}^{'}$ in this paragraph denotes differentiation by new
variable $y$ of extra dimension.
Similarly, we read the Kretschmann invariant from the metric (\ref{metric1})
\begin{equation}\label{KrB}
R^{ABCD}R_{ABCD}=\frac{p+1}{8B^{2}}\left(pB^{'4}+8B^{''2}\right).
\end{equation}

In subsequent sections, we shall discuss Randall-Sundrum type brane world
by use of the prepared building blocks.

\setcounter{equation}{0}
\section{Geometry of Randall-Sundrum Brane World}

\subsection{Pure anti-de Sitter spacetime}

\begin{figure}[ht]
\centerline{\scalebox{1}{\includegraphics{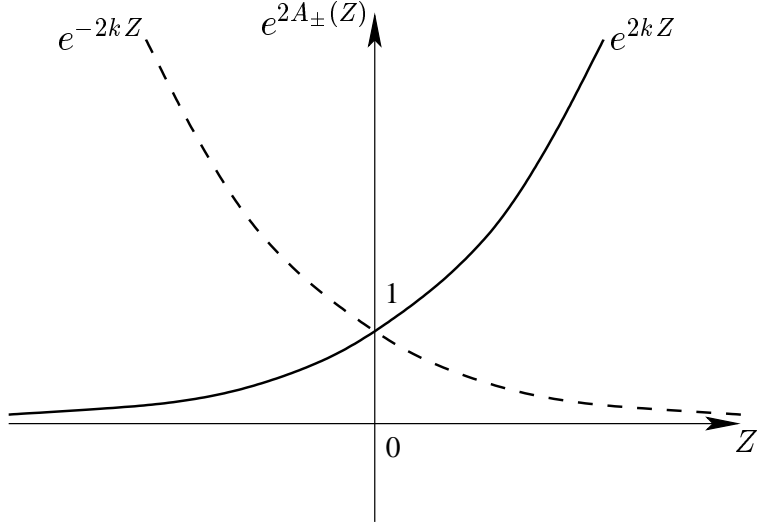}}}
\caption{The metric function of anti-de Sitter spacetime
$e^{2A_{\pm}(Z)}=e^{\pm 2kZ}$.}
\label{rsrfig1}
\end{figure}

When the bulk is filled only with negative vacuum energy $\Lambda<0$
without other matters $S_{\rm matter}=0$ so that $T_{AB}=0$, then
the Einstein equations (\ref{sezi})$\sim$(\ref{sezt}) are
\begin{equation}\label{u5s}
A^{''} = 0~ {\rm{and}} ~A^{'2} = - \frac{2\Lambda}{p(p+1)}.
\end{equation}
Notice that $A(Z)$ can have a real solution only when $\Lambda$ is nonpositive.
General solution of Eq.$~(\ref{u5s})$ is given by
\begin{equation}\label{apm}
A_{\pm}(Z)=\pm\sqrt{\frac{2|\Lambda|}{p(p+1)}}Z+A_{0},
\end{equation}
where the integration constant $A_{0}$ can be removed by rescaling of
the spacetime variables of $p$-brane, i.e., $dx^{\mu} \rightarrow
d\bar{x}^{\mu} = e^{A_{0}}dx^{\mu}$. The resultant metric is
\begin{equation}\label{uls}
ds^{2} =  e^{\pm 2kZ}
\eta_{\mu\nu} d\bar{x}^{\mu} d\bar{x}^{\nu} - dZ^{2},
\end{equation}
where $k = \sqrt{2|\Lambda| / p(p+1)}$ and a schematic shape
of the metric $e^{2A(Z)}$ is shown in Fig.~\ref{rsrfig1}.
Since the metric function $e^{2A_{\pm}}$ vanishes or is divergent at spatial
infinity $Z=\mp \infty$ respectively, there exists coordinate
singularity at those
points. Despite of the coordinate singularity, the spacetime is
physical-singularity-free everywhere as expected
\begin{equation}\label{adskret}
R^{ABCD}R_{ABCD}=\frac{8(p+2)}{p^{2}(p+1)}|\Lambda|^{2}.
\end{equation}

As mentioned in the previous section, the warp coordinate system
(\ref{uls}) is unusual
to depict geometry of the anti-de Sitter spacetime. A coordinate
transformation to the metric (\ref{metric1}) via $dZ=\alpha
dy/y$ leads to
\begin{eqnarray}
ds^{2} &=& e^{\pm\sqrt{\frac{8|\Lambda|}{p(p+1)}}\ln{\left(
\frac{y}{y_0}\right)^{\alpha}}}
\eta_{\mu\nu}d\bar{x}^{\mu}d\bar{x}^{\nu}
-\alpha^{2}\frac{dy^{2}}{y^{2}}\label{ads1}\\
&=& y^{\pm\alpha\sqrt{\frac{8|\Lambda|}{p(p+1)}}}\eta_{\mu\nu}
d\tilde{x}^{\mu}d\tilde{x}^{\nu} -\alpha^{2}\frac{dy^{2}}{y^{2}}\\
&=&y^{2}\eta_{\mu\nu}d\tilde{x}^{\mu}d\tilde{x}^{\nu}
-\alpha^{2}\frac{dy^{2}}{y^{2}}\label{ads3}\\
&=&\frac{p(p+1)}{2|\Lambda|}\left(y^{2}\eta_{\mu\nu}dx^{\mu}
dx^{\nu} -\frac{dy^{2}}{y^{2}}\right). \label{ads4}
\end{eqnarray}
The integration constant $\ln y_{0}$ in the first line (\ref{ads1}) was
eliminated by rescaling of the spacetime variables of $p$-brane, i.e.,
$d\tilde{x}^{\mu} = e^{\mp\alpha\sqrt{8|\Lambda|/p(p+1)}
\ln y_{0}}d\bar{x}^{\mu}$. The third line (\ref{ads3})
was obtained by a choice of $\alpha$ as $\pm 2\sqrt{\frac{p(p+1)}{8|\Lambda|}}$.
A rescaling of a coordinate $dx^{\mu}=d\tilde{x}^{\mu}/\alpha$
leads to the line (\ref{ads4}).
Because of the coordinate transformation $Z=\ln{y^{\alpha}}$, $Z=\infty$
corresponds to $y=0$ when $\alpha<0$ (or $y=\infty$ when $\alpha>0$) and
$Z=0$ to $y=1$ so that the spacetime described by the
coordinate system (\ref{stamet})
does not represent entire
anti-de Sitter spacetime but a patch of it
as is obvious from the coordinate transformation, $Z=\ln{y^{\alpha}}$. Now the
developed coordinate singularity is found at both $y=0$ and $y=\infty$ in
the metric (\ref{ads4}). Of course, the Kretschmann invariant
(\ref{KrB}) is independent of the choice of specific form of the metric
so that it is the same as Eq.~(\ref{adskret}). An intriguing observation
is that the coordinate singularity at $y=0$ can also be understood as
a horizon with zero radius limit (or equivalently zero mass limit) of
black $p$-brane.

\subsection{Randall-Sundrum brane world II}
\begin{figure}[ht]
\centerline{\scalebox{1}{\includegraphics{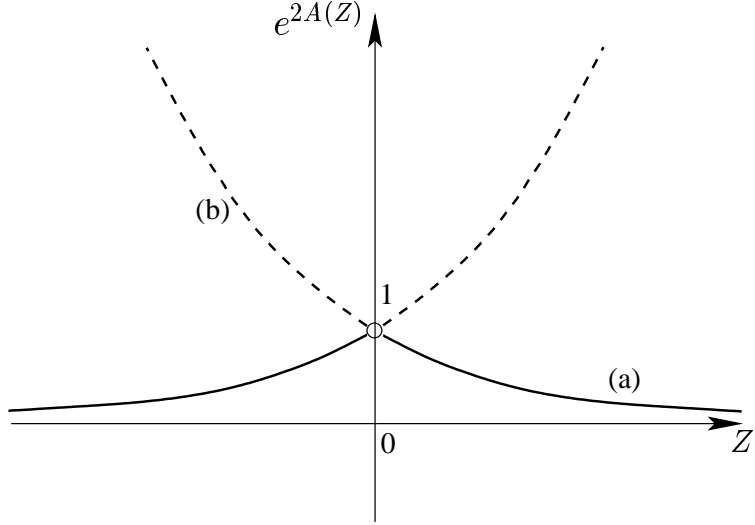}}}
\caption{Continuous Z${}_{2}$-symmetric anti-de Sitter space : (a)
$e^{2A_{\rm II}(Z)}=e^{-2k|Z|}$(solid line), (b) $e^{2A(Z)}=e^{2k|Z|}$
(dashed line).}
\label{rsrfig2}
\end{figure}

\begin{figure}[ht]
\centerline{\scalebox{1}{\includegraphics{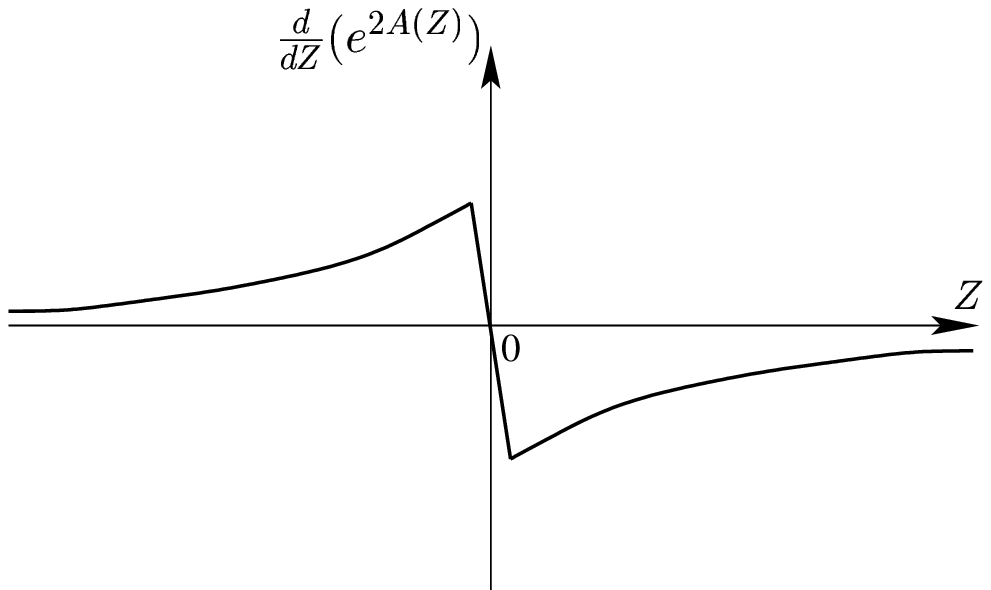}}}
\centerline{\scalebox{1}{\includegraphics{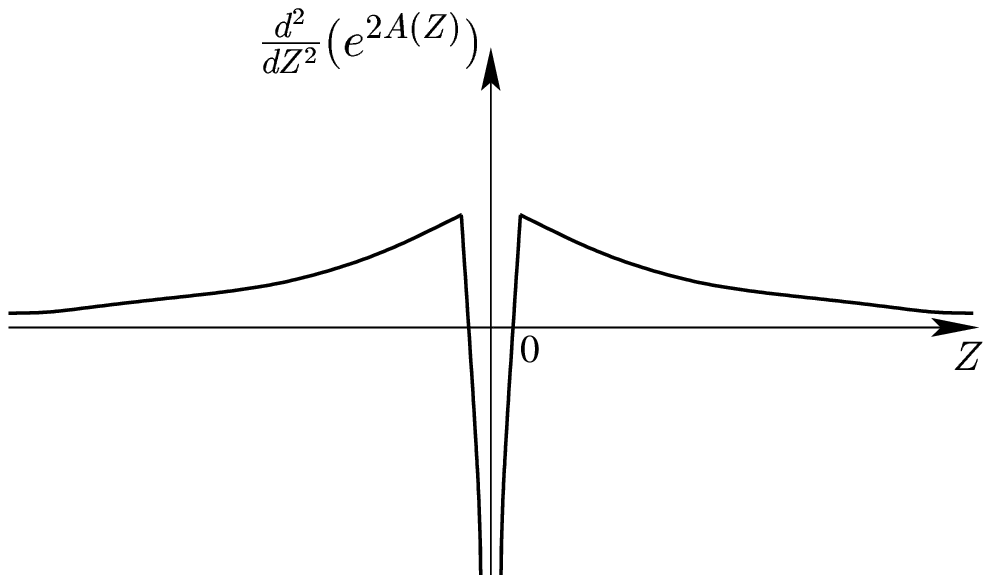}}}
\caption{(a) First derivative of $e^{2A_{\rm II}(Z)}$, (b) second
derivative of $e^{2A_{\rm II}(Z)}.$}
\label{rsrfig3}
\end{figure}

When we want to use the obtained solutions ~(\ref{apm}) for compactification
of the extra-dimension $\{Z\}$, the metric function should necessarily be
single-valued even at infinity $Z=\pm \infty$. A natural method is to urge
a reflection symmetry (${\rm Z}_{2}$-symmetry)
to $Z$-coordinate so that we can have two continuous solutions in
Fig.~\ref{rsrfig2} by patching two solutions (\ref{apm}) at the origin $Z=0$.
 Since we are not interested in exponentially-blowing up solution in
Fig.~\ref{rsrfig2}-(b), we consider only the
exponentially-decreasing warp factor in Fig.~\ref{rsrfig2}-(a)
from now on. Though it is continuous, it does not satisfy the
Einstein equations (\ref{u5s}) at the origin $Z=0$ as far as we
do not assume a singular matter configuration at that point. The
curve of the first derivative of $A(Z)$ is given by the step
functions
\begin{eqnarray}
A^{'}_{\rm II}(Z)=-k\left[ \theta(Z)-\theta(-Z)\right],\label{apr}
\end{eqnarray}
and thereby that of second derivative is nothing but a delta function
instead of zero as in Eq.~(\ref{u5s})
\begin{eqnarray}
A^{''}_{\rm II}(Z)= -2k\delta(Z).\label{aii}
\end{eqnarray}
Schematic shapes of first and second derivatives of the warp factor
$e^{2A_{\rm II}(Z)}$ are
shown in Fig.~\ref{rsrfig3}.

An appropriate interpretation of the delta function in Eq.~(\ref{aii})
is to regard it as a matter source confined on the $p$-brane at $Z=0$.
Eq.~(\ref{seti}) tells us that $T^{t}_{\;t}=T^{i}_{\;i}$ for any static
metric.
Substituting Eq.~(\ref{aii}) into one of the Einstein equations
(\ref{sezt}), we obtain
\begin{equation}
T^{t}_{{\rm II}\,t}-T^{Z}_{{\rm II}\,Z}=\frac{pkM^{p}_{\ast}}{4\pi}\delta(Z).
\end{equation}
Insertion of Eq.~(\ref{apr}) into the $ZZ$-component of the Einstein
equation
(\ref{sezi}) provides vanishing $ZZ$-component of the energy-momentum tensor
\begin{equation}
T^{Z}_{{\rm II}\,Z}=\frac{M^{p}_{\ast}|\Lambda|}{8\pi}\left\{
[\theta(Z)-\theta(-Z)]^{2}-1\right\}=0.
\end{equation}
Therefore, we have
\begin{equation}\label{tttt}
T^{t}_{{\rm II}\,t}=T^{i}_{{\rm II}\,i}=\frac{M^{p}_{\ast}}{8\pi}
2pk\delta(Z),~~~T^{Z}_{{\rm II}\,Z}=0,
\end{equation}
and corresponding covariant form of it is
\begin{eqnarray}\label{em20}
T^{\;A}_{{\rm II}\,B}=\frac{M^{p}_{\ast}}{8\pi}2pk\delta(Z)
\delta^{\mu}_{\;\nu}\delta^{A}_{\;\mu}\delta_{\;B}^{\nu}.
\end{eqnarray}

This is the result we expected, that is, the delta function
source in Eq.~(\ref{aii}) is indeed a constant matter density on
the $p$-brane at the origin. Signature of the energy-momentum
tensor (\ref{em20}) implies the positiveness of the $p$-brane
tension. When the matter is confined on a specific $p$-brane, the
metric fluctuation to the radial direction vanishes, i.e. $\delta
g^{ZZ}=0$. Therefore the energy-momentum tensor from the sources
confined on the $p$-brane becomes
\begin{equation}
T^{\rm brane}_{AB}\equiv \left. \frac{2}{\sqrt{g_{D}}}\frac{\delta
S^{\rm brane}_{\rm matter}}{\delta g^{AB}}\right|_{\delta
g^{ZZ}=0} .
\end{equation}

An appropriate form of matter action is written by use of
Eq.~(\ref{emtb}) such as
\begin{eqnarray}\label{brm1}
S_{\rm II}= \frac{M^{p}_{\ast}}{8\pi}\int d^{p+1}x
\int^{\infty}_{-\infty}dZ\sqrt{g_{D}}\, 2pk\delta(Z).
\end{eqnarray}
Note that the above junction condition
at the $p$-brane is nothing but a fine-tuning
condition since all the contents of the matter action (\ref{brm1}) should be
determined by the quantities of the bulk, specifically by the fundamental
scale of the bulk theory $M_{\ast}$ and the bulk cosmological constant
$\Lambda$. Since there is no constant density term in the $p$-brane action,
the effective cosmological constant on the $p$-brane (or our universe)
vanishes.

The resultant metric of Randall-Sundrum brane world model II~\cite{RS2} is
\begin{equation}\label{resmet}
ds^{2} =  e^{-2k|Z|}
\eta_{\mu\nu} d\bar{x}^{\mu} d\bar{x}^{\nu} - dZ^{2}.
\end{equation}
Once we transform it to the Schwarzschild-type coordinates (\ref{metric1})
, we can easily find coordinate singularities at both infinity,
$Z=\pm \infty$. Since we added the matter on the $p$-brane as a delta function
source, the Kretschmann invariant contains a delta function like physical
singularity at $Z=0$
\begin{equation}
R^{ABCD}R_{ABCD}=\frac{8}{p}|\Lambda|\left[\frac{|\Lambda|}{p+1}+
\left(\sqrt{\frac{2|\Lambda|}{p(p+1)}}-2\delta (Z)\right)^{2}\right].
\end{equation}

\subsection{Randall-Sundrum brane world I}
Suppose that the coordinate of the extra-dimension $Z$ is really compact in
Randall-Sundrum brane world model I, different from the
previous Randall-Sundrum brane world II with $-\infty \le Z \le \infty$.
An appropriate method from the brane world II to I is attained by forcing
periodicity to the coordinate of the extra-dimension $Z$ in addition to the
Z${}_{2}$-symmetry as shown in
Fig.~\ref{rsrfig4}. It is exactly an orbifold compactification by
${\bf S}^{1}/Z_{2}$ and thereby physics of our interest lives in a
compact region $\{ Z | [0,r_{\rm c}\pi]\}$. To achieve this geometry by adding
matters on the branes at both $Z=0$ and $Z=r_{\rm c}\pi$,
we already learned that two delta function sources should be taken into
account at both $Z=0$ and $Z=r_{\rm c}\pi$, similar to the action
(\ref{brm1}) :
\begin{figure}[ht]
\centerline{\scalebox{1}{\includegraphics{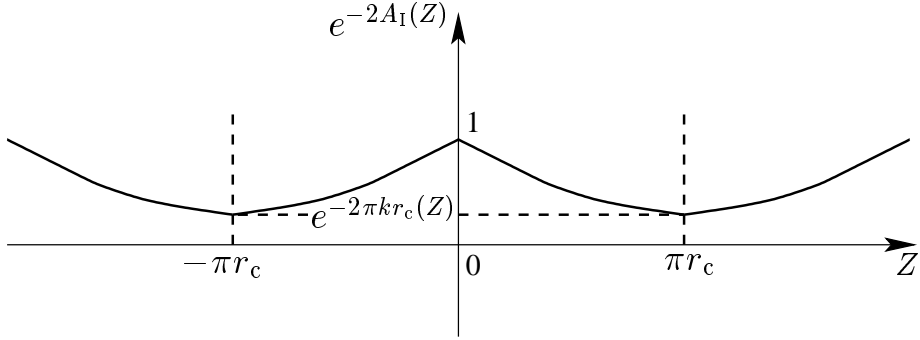}}}
\caption{Metric function $e^{-2A_{\rm I}(Z)}$ for Randall-Sundrum
compactification I. There is a hidden brane matter at $Z=0$, and our world
in electroweak scale is located at $Z=r_{\rm c}\pi$.}
\label{rsrfig4}
\end{figure}
\begin{eqnarray}
S_{\rm I} &\equiv& S_{\rm I}|_{Z=0}+S_{\rm I}|_{Z=r_{\rm c}\pi} \nonumber\\
&=& \frac{M^{p}_{\ast}}{8\pi}\int d^{p+1} x
\int^{r_{\rm c}\pi}_{-r_{\rm c}\pi} dZ\sqrt{g_{D}}\, 2pk[\delta(Z)-\delta(Z -
r_{\rm c}\pi)].
\end{eqnarray}
Then the corresponding energy-momentum tensor restricted on both $p$-branes
is computed by the formula (\ref{emtb})
\begin{eqnarray}
T^{A}_{{\rm I}\; B}= \frac{M^{p}_{\ast}}{8\pi}2pk[\delta(Z)-\delta(Z -
r_{\rm c}\pi)]
\delta^{\mu}_{\;\nu}\delta^{A}_{\;\mu}\delta_{\;B}^{\nu},
\end{eqnarray}
and the Einstein equations (\ref{sezi})$\sim$(\ref{sezt}) become
\begin{eqnarray}
A^{'}_{\rm I}(Z)&=&-2k\left[\theta(Z)-\theta(-Z)\right]
-2k\left[\theta(Z-r_{\rm c}\pi)-\theta(-Z+r_{\rm c}\pi)\right],
\label{wein4} \\
A^{''}_{\rm I}(Z)&=&-2k[\delta(Z)-\delta(Z-r_{\rm c}\pi)].
\label{wein3}
\end{eqnarray}
Note that the brane at $Z=0$ has positive tension but the other brane
at $Z=r_{\rm c}\pi$ has negative tension.
The metric of the Randall-Sundrum brane world I~\cite{RS1} is expressed by
\begin{equation}
ds^{2}=e^{-2kr_{\rm c}\varphi}\eta_{\mu\nu}dx^{\mu}dx^{\nu}-r_{\rm c}^{2}
d\varphi^{2},~~~(0\le \varphi \le \pi).
\end{equation}
It is free from coordinate singularity but involves
physical singularity at both patched boundaries ($Z=0$ and $Z=r_{\rm c}\pi$):
\begin{equation}
R_{ABCD}R^{ABCD}=\frac{8}{p}|\Lambda|\left\{\frac{|\Lambda|}{p+1}+
\left[\sqrt{\frac{2|\Lambda|}{p(p+1)}}-2\left(\delta (Z)-\delta
(Z-r_{\rm c}\pi)\right)
\right]^{2}\right\}.
\end{equation}

Again, we encounter the fine-tuning conditions: One is the fine-tuning
that
the brane matter action is completely determined by the bulk negative
cosmological constant $\Lambda$ and the fundamental scale of the bulk theory
$M_{\ast}$, and the other is the fine-tuning that the magnitudes of both brane
matter actions, $S_{\rm I}|_{Z=0}$ and $S_{\rm I}|_{Z=r_{\rm c}\pi}$,
are exactly the same each other but have the opposite sign:
\begin{equation}\label{ftc}
\frac{S_{\rm I}|_{Z=0}}{V_{p+1}}=-\frac{S_{\rm I}|_{Z=r_{\rm c}\pi}}{V_{p+1}}
=\frac{M^{p}_{\ast}}{8\pi},
\end{equation}
where the spacetime volume of each $p$-brane is denoted by $V_{p+1}=
\int d^{p+1}x$. Therefore, the $p$-brane at the origin has the positive
tension
but the other brane at $Z=r_{\rm c}\pi$ the negative tension. Note that
the
effective cosmological constant vanishes on the $p$-branes at both boundaries,
$Z=0$ and $Z=r_{\rm c}\pi$.

\setcounter{equation}{0}
\section{Physical Implication of Randall-Sundrum Brane World}
In this section we discuss two main features of Randall-Sundrum brane
world models.
In the model II with single brane, gravitational fluctuations on the brane
reproduce the Newtonian gravity from normalizable zero mode.
In the model I with two branes, the gauge hierarchy problem
can be treated in a much milder form without assuming supersymmetry.

\subsection{Newtonian gravity from model II}


In the subsection 3.2, we discussed the Randall-Sundrum brane
world model II which can be defined whthin $-\infty \le Z \le
\infty$. The summary for this RS II model is described by the
metric in Eq.~(\ref{resmet}). The aim of this subsection is to
determine whether the spectrum of general linerized tensor
fluctuations $H_{\mu \nu}$ is consistent with 4-dimensional
experimental gravity. To do so, let us consider the small
gravitational fluctuations $\delta g_{AB}$ on the given background
metric $g_{AB}$.
\begin{equation}\label{dmet}
ds^{2} = (g_{AB}+\delta g_{AB})dx^{A}dx^{B}.
\end{equation}
In present RS model II, we restrict the small fluctuations
$\delta g_{AB}$ to the metric $h_{\mu\nu}(x,Z)$ of 4-dimensional 
world on the 3-brane. The metric in Eq. (\ref{dmet})
becomes
\begin{equation}\label{fmet}
ds^{2} =  [e^{-2k|Z|}\eta_{\mu\nu} +h_{\mu\nu}(x,Z)] dx^{\mu}
dx^{\nu} - dZ^{2}\equiv H_{\mu \nu}dx^{\mu} dx^{\nu} - dZ^{2},
\end{equation}
where $H_{\mu \nu}$ stands for the linearized tensor fluctuations. 
Substituting the metric (\ref{fmet}) into the
Einstein tensor $G_{AB}$ with the help of transverse-traceless
gauge where
\begin{equation}\label{rsga}
\partial^{\mu}h_{\mu\nu}=0,\qquad h^{\mu}_{\; \mu}=0,
\end{equation}
we can easily see that the small fluctuations in the Einstein tensor $G_{AB}$
have the nonvanishing components only on the 3-brane as
\begin{eqnarray}
&&\hspace{-15mm}\delta\left( G_{\mu\nu}=-\frac{8\pi}{M^{p}_{\ast}}T_{\mu\nu}
+\Lambda g_{\mu\nu} \right) \nonumber\\
&&\Longrightarrow\hspace{5mm}
\left[\frac{1}{2}\left(U^{-1}\eta^{\rho\sigma}\partial_{\rho}\partial_{\sigma}
-\partial_{Z}^{2} \right)+V(Z)\right]h_{\mu\nu}=0,
\label{lieq}
\end{eqnarray}
where
\begin{eqnarray}
V(Z)&=&2\left(\frac{U''}{U}\right)+6k\delta(Z)-6k^{2},\\
U(Z)&=&e^{-2k|Z|}.
\end{eqnarray}
See Appendix C for detailed derivation of Eq.~(\ref{lieq}).

To understand all modes that appear in 4-dimensional
effective theory, we perform a KK reduction down to
four dimensions. To do so, let us summarize the obtained linearized
equations for the small fluctuations
\begin{equation}\label{fili}
\left[\frac{1}{2}\left(e^{2k|Z|}\eta^{\rho\sigma}\partial_{\rho}
\partial_{\sigma} -\partial_{Z}^{2} \right)-2k\delta(Z)+2k^{2}
\right]h_{\mu\nu}(x^{\alpha},Z)=0.
\end{equation}
Since nontrivial potential part depends only on the
5th-coordinate $Z$, we can easily apply the separation of
variables to this linear equation. To be specific, inserting
\begin{equation}
h_{\mu\nu}(x^{\rho},Z)=\psi(Z) \Phi(x^{\rho})
\end{equation}
into Eq.~(\ref{fili}), we have
\begin{eqnarray}
&&\eta^{\mu\nu}\partial_{\mu}\partial_{\nu}\Phi(x^{\rho})=-m^{2}
\Phi(x^{\rho}),\qquad (m^{2}\ge 0),
\label{Newe}\\
&&\left[-\frac{m^{2}}{2}e^{2k|Z|} -\frac{1}{2}\partial_{Z}^{2}
-2k\delta(Z)+2k^{2} \right]\psi(Z)=0,
\label{sieq}
\end{eqnarray}
where $m$ is the 4-dimensional mass of the KK excitation.

By making a change of variable as follows
\begin{eqnarray}
w&=&\frac{{\rm sgn}(Z)}{k}(e^{k|Z|}-1),\qquad ({\rm sgn}(Z)\equiv Z/|Z|
=\theta(Z)-\theta(-Z)),\\
\hat{\psi}(w)&=&e^{k|Z|/2}\psi(Z),
\end{eqnarray}
we rewrite Eq.~(\ref{sieq}) in a simpler form
\begin{equation}\label{pleq}
-\frac{1}{2}\frac{d^{2}\hat{\psi}}{dw^{2}}
+\hat{V}(w)\hat{\psi}=\frac{m^{2}}{2}\hat{\psi},
\end{equation}
where
\begin{equation}\label{poeq}
\hat{V}(w)=\frac{15k^{2}}{8(k|w|+1)^{2}}-\frac{3}{2}k\delta(w).
\end{equation}
Here we used  $\delta(w)=\delta(Z)=\frac{1}{2}\frac{d}{dZ}{\rm
sgn}(Z)$ and see Fig.~\ref{rsrfig3}-(b) for the volcano-type
potential $\hat{V}$. Since we have an explicit form of the KK potential
~(\ref{poeq}),
we will discuss the properties of continuum modes $m$ in the end of this
subsection. Before doing
so, however, we would like to give the discussions on the case of
zero mode, $m^2 =0$, in Eq.~(\ref{Newe}).

In the static frame with the rotational symmetry on the 3-brane (or 
our universe), Eq.~(\ref{Newe}) with $m^2 =0$ is reduced to
the well-known Laplace equation
\begin{equation}\label{laeq}
\frac{1}{r^{2}}\frac{d}{dr}\left[r^{2}\frac{d\Phi(r)}{dr}\right]=0.
\end{equation}
Except for the source point at the origin $r=0$, the Newtonian potential
\begin{equation}\label{New}
\Phi(r)=-\frac{A}{r}
\end{equation}
satisfies Eq.~(\ref{laeq}). Here we set $\Phi(\infty)=0$ and
$A=G_{{\rm N}}m_{1}m_{2}$ in order to match Newtonian gravity
between two particles of mass $m_{1}$ and $m_{2}$ on our brane at $Z=0$.
Now that we have Newtonian gravitational potential on the 3-brane, we
solve $\psi(Z)$ in the extra dimension. When $m^{2}=0$, we directly
deal with Eq.~(\ref{sieq}) given by
\begin{equation}\label{moeq}
\frac{d^{2}\psi}{dZ^{2}}=\left[4k^{2}-4k\delta(Z)\right]\psi(Z).
\end{equation}
For $Z\ne 0$, we have
\begin{equation}
\psi(Z)=\psi_{0}e^{-2k|Z|},
\end{equation}
which satisfies the boundary condition obtained by integration of
Eq.~(\ref{moeq}) for $-\varepsilon \le Z \le \varepsilon$ for
infinitesimal $\varepsilon$
\begin{equation}
4k\psi(0)=\lim_{\varepsilon\rightarrow 0}\left(
\left.\frac{d\psi}{dZ}\right|_{\varepsilon}
-\left.\frac{d\psi}{dZ}\right|_{-\varepsilon}\right).
\end{equation}
Normalization condition $\int_{-\infty}^{\infty}dZ|\psi(Z)|^{2}=1$
fixes the overall constant $\psi_{0}$ as
\begin{equation}
\psi(Z)=\sqrt{2k}e^{-2k|Z|}.
\end{equation}

With the explicit form of the KK potential~(\ref{poeq}), we can understand the
properties of KK modes of $m^2 \not= 0$. Since the KK potential
falls off to zero as $|Z| \rightarrow
\infty$, the continuum KK states with no gap exist for all
possible $m^2 > 0$ and then the proper measure is simply
$dm$. For the detailed discussion on the proper measure through the
Bessel function representation for the solution of
Eq.~(\ref{pleq}) refer to Ref.~\cite{RS2}.

With the KK spectrum of the effective 4-dimensional theory,
let us compute the gravitational potential $\Phi(r)$ between two particles
of mass $m_1$ and $m_2$ on our brane at $|Z|=0$, which is the static potential
generated by exchange of the zero-mode and continuum KK modes
propargators;
\begin{equation}\label{phia}
\Phi
(r) \approx G_{\rm N} \frac{m_1 m_2}{r} + \frac{G_{\rm N}}{k} m_1 m_2
\int_0^{\infty} dm\frac{m}{k}
\frac{e^{-mr}}{r}.
\end{equation}
There is a Yukawa potential in the correction term, 
and an extra factor of $m/k$
comes from the continuum wave functions $\psi(Z)$ for $m^2 \not= 0
$ at $Z=0$. The coupling $G_{\rm N} / k$ is nothing but the fundamental
coupling of gravity, $1 / M_{*}^3$. By performing the
integration over $r$ in Eq.~(\ref{phia}), we have a next order correction
of $O(1/r^{3})$ to the Newtonian potential
\begin{equation}
\Phi(r) \sim G_{\rm N} \frac{m_1 m_2}{r}\left(1 + \frac{1}{k^2 r^2}\right).
\end{equation}
This is the reason why the RS II model produces an effective 4-dimensional
theory of gravity: the leading term is given by the usual Newtonian
potential and a continuos 
KK modes generate a correction term. Note that the radion
can be an additional source of $O(1/r^{3})$ contribution~\cite{GW}.


\subsection{Gauge hierarchy from model I}
As we explained briefly in the introduction, the gauge hierarchy problem is
a notorious fine tuning problem in particle phenomenology of which
the basic language is quantum field theory. So the readers unfamiliar to
field theories may skip this subsection.

Let us assume that we live on the $p$-brane at $Z=r_{\rm c}\pi$ and try a
dimensional reduction of the Einstein gravity from the $D=p+2$-dimensional
gravity to $p+1$-dimensional gravity on the $p$-brane at $Z=r_{\rm c}\pi$.
Then we have
\begin{eqnarray}
S_{{\rm EH}\, D}&=&-\frac{M_{\ast}^{p}}{16\pi}\int d^{D}x\sqrt{|g_{D}|}\,R
\label{first}\\
&=&-\frac{M_{\ast}^{p}}{16\pi}\int d^{p+1}x \sqrt{|\det g_{\mu\nu}|}
\int^{r_{\rm c}\pi}_{-r_{\rm c}\pi}dZe^{-(p-1)k|Z|}
(R_{p+1}+\cdots)
\label{second}\\
&=&-\frac{M_{\ast}^{p}}{16k\pi}\left[1-e^{-(p-1)kr_{\rm c}\pi}\right]
\int d^{p+1}x \sqrt{|\det g_{\mu\nu}|}\, (R_{p+1}+\cdots)
\label{5dp}\\
&\equiv&-\frac{M_{\rm Planck}^{2}}{16\pi}\int d^{p+1}x \sqrt{|\det g_{\mu\nu}|}
\,(R_{p+1}+\cdots) \\
&=&S_{{\rm EH}\, p+1}+\cdots .
\label{4dp}
\end{eqnarray}
We used $g_{D}=e^{-2(p+1)k|Z|}\det g_{\mu\nu}$ and $R=e^{2k|Z|}g^{\mu\nu}
R_{\mu\nu}+\cdots=e^{2k|Z|}R_{p+1}+\cdots$ when we calculated the second line
(\ref{second}) from the first line (\ref{first}).
By comparing the third line (\ref{5dp}) with the fourth line (\ref{4dp}), we
obtain a relation for $3$-brane among three scales
$M_{\rm Planck}$, $M_{\ast}$, $|\Lambda|$ $(p=3)$:
\begin{eqnarray}\label{relm}
M_{\rm Planck}^{2}=\sqrt{\frac{p(p+1)}{2|\Lambda|}}\left[1-\exp\left(
-\sqrt{\frac{8|\Lambda|}{p(p+1)}}r_{\rm c}\pi\right)\right]M_{\ast}^{p=3}.
\end{eqnarray}
A natural choice for the bulk theory is to bring up almost the same scales
for two bulk mass scales, i.e., $M_{\ast}\approx \sqrt{|\Lambda|}$. Suppose
that the exponential factor in the relation (\ref{relm}) is negligible to the
unity, which means $r_{\rm c}$ is slightly larger than $1/\sqrt{|\Lambda|}$.
Then we reach
\begin{equation}
M_{\rm Planck}\approx M_{\ast}\approx \sqrt{|\Lambda|}.
\end{equation}

A striking character of this Randall-Sundrum compactification I is that
it provides an explanation for gauge hierarchy problem that {\it why is so
large the mass gap between the Planck scale $M_{\rm Planck}\sim 10^{19}
{\rm GeV}\sim 10^{-38}~M_{\odot}$ 
and the
electroweak scale $M_{\rm EW}\sim 10^{3}
{\rm GeV}\sim 10^{-54}$~$M_{\odot}$} without assuming supersymmetry or others.
As a representative example, let us
consider a massive neutral scalar field $H$ which lives on our $3$-brane at
$Z=r_{\rm c}\pi$ :
\begin{eqnarray}
S_{\rm scalar}&=&\int^{r_{\rm c}\pi}_{-r_{\rm c}\pi}dZ
\delta(Z-r_{\rm c}\pi)
\int d^{4}x\sqrt{g_{5}} \Biggl[\frac{1}{2}g^{AB}
\partial_A H \partial_B H -\frac{1}{2}M^{2}_{{\rm Planck}}H^2 \Biggr]
\nonumber \\
&=&  \int^{r_{\rm c}\pi}_{-r_{\rm c}\pi}dZ e^{-4k|Z|}\delta(Z-r_{\rm c}\pi)\int
d^{4}x
\sqrt{-\hat{g}_{4}}\nonumber\\
&&\hspace{10mm}\times\Biggl[\frac{1}{2}
e^{2k|Z|}\hat{g}^{\mu\nu}\partial_{\mu}H\partial_{\nu}H-\frac{1}{2}
M^{2}_{{\rm Planck}} H^2 - \frac{1}{2}\hat{g}^{ZZ}\left(\partial_Z H\right)^
{2} \Biggr]\nonumber\\
&=& e^{-2r_{\rm c}\pi k} \int d^{4}x\sqrt{-\hat{g}_{4}}
\Biggl[\frac{1}{2}\hat{g}^{\mu\nu}
\partial_{\mu}H\partial_{\nu}H - \frac{1}{2}(e^{-r_{\rm c}\pi k}M_{{\rm
Planck}})
^{2}H^{2} \Biggr]\\
&=&  e^{-2r_{\rm c}\pi k}\int
d^{4}x\sqrt{-\hat{g}_{4}}\Biggl[\frac{1}{2}\hat{g}^{\mu\nu}
\partial_{\mu}H\partial_{\nu}H - \frac{1}{2}M_{{\rm EW}}^{2}H^{2}\Biggr],
\end{eqnarray}
where $ds^{2}=g_{AB}dx^A dx^B=e^{-2k|Z|}\hat{g}_{\mu\nu}dx^{\mu}dx^{\nu}-dZ^2$.
The last two lines give us a relation:
\begin{eqnarray}\label{hier}
\frac{M_{\rm EW}}{M_{\rm Planck}}=
\exp\left(-\sqrt{\frac{2|\Lambda|}{p(p+1)}}r_{\rm c}\pi\right)\sim 10^{-16}.
\end{eqnarray}
Therefore, the radius $r_{\rm c}$ of compactified extra dimension
of the Randall-Sundrum
brane world model I is determined nearly by the Planck scale :
\begin{equation}
\frac{1}{r_{\rm c}}\sim \frac{\pi}{16\sqrt{6}\ln 10}\sqrt{|\Lambda|}
\sim \frac{M_{\rm Planck}}{30}.
\end{equation}
All the scales such as the fundamental scale of the bulk $M_{\ast}$, the
bulk cosmological constant $\sqrt{|\Lambda|}$, the inverse size of the
compactification $1/r_{\rm c}$, are almost the Planck scales $M_{\rm Planck}
\sim 10^{19}$ GeV
together. The masses of matter particles on our visible brane at
$Z=r_{\rm c}\pi$
are in electroweak scale $M_{\rm EW}\sim 10^{3}$ GeV,
however those on the hidden brane at $Z=0$ in the Planck scale. Though the
gauge hierarchy problem seems to be solved, it is actually not because a
fine-tuning condition was urged in Eq.~(\ref{ftc}). However, it becomes much
milder than that before.

How about a massive gauge field $A_{\mu}$ which lives on our
$3$-brane at $Z=r_{\rm c}\pi$? We have
\begin{eqnarray}
S_{\rm gauge}&=&\int^{r_{\rm c}\pi}_{-r_{\rm c}\pi}dZ
\delta(Z-r_{\rm c}\pi) \int d^{4}x\sqrt{g_{5}}
\Biggl[-\frac{1}{4}g^{AC}g^{BD}F_{AB}F_{CD}
+ M^{2}_{\rm Planck}g^{AB}A_{A}A_{B} \Biggr] \nonumber \\
&=& \int^{r_{\rm c}\pi}_{-r_{\rm c}\pi}dZ
e^{-4k|Z|}\delta(Z-r_{\rm c}\pi)\int d^{4}x
\sqrt{-\hat{g}_{4}}\nonumber\\
&&\hspace{2mm}\times\Biggl[-\frac{1}{4}e^{4k|Z|}\hat{g}_{\mu\rho}
\hat{g}_{\nu\lambda} F_{\mu\nu}F_{\rho\lambda}+M^{2}_{\rm
Planck}e^{2k|Z|}\hat{g}^{\mu\nu}A_{\mu}A_{\nu}\Biggr]\nonumber\\
&=& \int d^{4}x\sqrt{-\hat{g}_{4}}
\Biggl[-\frac{1}{4}\hat{g}^{\mu\rho}\hat{g}^{\nu\sigma}
F_{\mu\nu}F_{\rho\sigma}
+(e^{-r_{\rm c}\pi k}M_{\rm Planck})^{2}
\hat{g}^{\mu\nu}A_{\mu}A_{\nu}\Biggr] \label{gauge1}\\
&=& \int
d^{4}x\sqrt{-\hat{g}_{4}}\Biggl[-\frac{1}{4}\hat{g}^{\mu\rho}\hat{g}^{\nu\sigma}
F_{\mu\nu}F_{\rho\sigma} +m^{2}_{\rm
gauge}\hat{g}^{\mu\nu}A_{\mu}A_{\nu}\Biggr].\label{gauge2}
\end{eqnarray}
From Eq.~(\ref{gauge1}) and Eq.~(\ref{gauge2}) with Eq.~(\ref{hier}),
we read exactly the same mass hierarchy for the gauge field:
$m_{\rm gauge}=e^{-r_{\rm c}\pi k}M_{\rm Planck}=M_{\rm EM}$.
Therefore the gauge hierarchy can be interpreted by introducing
the massive gauge field similar to the case of the massive
neutral scalar field $H$.

Finally let us consider a fermionic field of which mass is provided
by spontaneous symmetry breaking and its Lagrangian is
\begin{equation}\label{yu1}
{\cal L}_{\rm fermion}=\bar{\Psi}\gamma^{A}\nabla_{A}\Psi+
g\phi\bar{\Psi}\Psi,
\end{equation}
where $g$ is the coupling constant of Yukawa interaction.
If we neglect the quantum
fluctuation $\delta\phi$ of $\phi$, i.e. $\phi \equiv
\left<\phi\right>+\delta \phi$, the Lagrangian (\ref{yu1}) becomes
\begin{equation}\label{yu2}
{\cal L}_{\rm fermion}=\bar{\Psi}\gamma^{A}\nabla_{A}\Psi +
g\left<\phi\right>\bar{\Psi}\Psi + \cdots ,
\end{equation}
where the second term is identified as mass term,
and we neglected the vertex term $g\delta\phi\bar{\Psi}\Psi$
because we are not
interested in quantum fluctuation.
Again the fermion lives on our 3-brane at $Z=r_{\rm c}\pi$, and then
the action is
\begin{equation}\label{yu3}
S_{\rm fermion}=\int^{r_{\rm c}\pi}_{r_{\rm c}\pi}dZ
\delta(Z-r_{\rm c}\pi) \int
d^{4}x\sqrt{g_{5}}\left[\bar{\Psi}\gamma^{a}e^{A}_{a}\nabla_{A}\Psi +
M_{\rm Planck} \bar{\Psi}\Psi\right],
\end{equation}
where $e^{A}_{a}$ is vielbein defined by
$g_{AB}=\eta_{ab}e^{a}_{A}e^{b}_{B}$ and
$M_{\rm Planck}=g\left<\phi\right>$ since the symmetry breaking scale
should coincide with the fundamental scale.
Subsequently, the action (\ref{yu3}) becomes
\begin{eqnarray}
S_{\rm fermion}&=&\int^{r_{\rm c}\pi}_{r_{\rm c}\pi}dZ
\delta(Z-r_{\rm c}\pi) \int
d^{4}x\sqrt{g_{5}}\Big[\bar{\Psi}\gamma^{a}e^{A}_{a}\nabla_{A}\Psi
+ M_{\rm Planck} \bar{\Psi}\Psi\Bigr] \nonumber\\
&=& \int^{r_{\rm c}\pi}_{-r_{\rm c}\pi}dZ
e^{-4k|Z|}\delta(Z-r_{\rm c}\pi)\int d^{4}x \sqrt{-\hat{g}_{4}}\nonumber\\
&&\hspace{7mm}\times\Bigl[e^{k|Z|}\bar{\Psi}
\gamma^{a}\hat{e}^{\mu}_{a}\nabla_{\mu}\Psi
- \bar{\Psi}\gamma^{a}\hat{e}^{Z}_{a}\nabla_{Z}\Psi +
M_{\rm Planck} \bar{\Psi}\Psi \Bigr]\nonumber\\
&=& e^{-3r_{\rm c}\pi k}\int d^{4}x\sqrt{-\hat{g}_{4}}\,
\Bigl[\bar{\Psi}\gamma^{a}\hat{e}^{\mu}_{a}\nabla_{\mu}\Psi +
(e^{-r_{\rm c}\pi k}M_{\rm Planck}) \bar{\Psi}\Psi\Bigr] \label{yu5}\\
&=& e^{-3r_{\rm c}\pi k}\int
d^{4}x\sqrt{-\hat{g}_{4}}\,\Bigl[\bar{\Psi}\gamma^{a}\hat{e}^{\mu}_{a}
\nabla_{\mu}\Psi + m_{\rm fermion} \bar{\Psi}\Psi \Bigr].
\label{yu6}
\end{eqnarray}
Once again we obtain the same mass hierarchy relation
$m_{\rm fermion}=e^{-r_{\rm c}\pi k}M_{\rm Planck}=M_{\rm EW}$
for the fermion from
Eq.~(\ref{yu5}) and Eq.~(\ref{yu6}) with the help of Eq.~(\ref{hier}).

In this subsection, we demonstrate how to understand the gauge hierarchy
problem in the context of Randall-Sundrum brane world model I.

\setcounter{equation}{0}
\section{Concluding Remarks}

In this review, we explained original idea of
Randall-Sundrum brane world models
I and II. RS I provided a geometrical resolution based on the warp factor
to make the gauge hierarchy problem much milder.
Though the bulk of RS II contains negative bulk cosmological constant,
its effect is cancelled by adjusting the 3-brane tension and then
Newtonian gravity is reproduced in weak gravity limit with subleading
KK modes
on the 3-brane identified as our universe.

Let us conclude by providing some
information on a several research topics in this field.
They include
the problem finding general form of RS solution~\cite{KK},
the stability of brane world model including radion~\cite{GW},
a variety of brane world models basically similar to RS models~\cite{GRS},
cosmological implication of RS model including reproduction of standard
cosmology~\cite{BDL},
construction of thick brane world particularly in terms of solitonic
object~\cite{Od},
finding supersymmetry in brane world~\cite{ABN},
RS model in the context of string theory~\cite{CCV},
brane world with extra dimensions more than one~\cite{CK},
implication to particle phenomenology~\cite{APR},
classical solutions which self-tune the cosmological constant~\cite{ADKS},
and CMB anisotropy study in brane world~\cite{KIS}.

\section*{Acknowledgments}
{We would like to thank GungWon Kang and Hang Bae Kim
for valuable discussions and comments.
This work is the result of
research activities (Astrophysical Research
Center for the Structure and Evolution of the Cosmos (ARCSEC))
supported by Korea Science $\&$ Engineering Foundation(Y.K.).}

\appendix

\setcounter{equation}{0}
\section{Units and Notations}
For convenience, we summarize the unit system and the various quantities
in this appendix. Our unit system is based on
$\hbar(\equiv h/2\pi)=c=1$. Since the light speed $c$ is set to be one,
mass of a particle $M$ and its rest energy $Mc^{2}$ have the same unit.
Since $c=3\times 10^{8}$m/sec and 1J=1kg$\,$m${}^{2}$/sec${}^{2}$
$\sim 10^{19}$eV, we have 1kg$\sim 10^{27}$GeV. Astronomical unit of mass is
expressed by solar mass $M_{\odot}\sim 2\times 10^{30}$kg.
Mass scales are given in Table 1.
\begin{center}
\renewcommand{\arraystretch}{1.7}
\begin{tabular}{|c |c cc|}\hline
mass & particle\hspace{8mm} & daily\hspace{8mm} & astronomy $\&$ \\
scale & physics\hspace{8mm} & life\hspace{8mm} & astrophysics \\ \hline
Planck & $M_{{\rm Planck}}\sim 10^{19}$GeV\hspace{8mm} & $\sim 10^{-8}$kg
\hspace{8mm} & $\sim 10^{-38}M_{\odot}$ \\
Electroweak & $M_{\rm EW}\sim 10^{3}$GeV\hspace{8mm} & $\sim 10^{-24}$kg
\hspace{8mm} & $\sim 10^{-54}M_{\odot}$ \\ \hline
\end{tabular}
\end{center}
Our basic conversion relation is
\begin{equation}
\hbar c\approx 2\times 10^{-16} {\rm GeV}\cdot {\rm m}.
\end{equation}
Therefore, uncertainty principle $\Delta E (c\Delta t)\sim \hbar c$ tells us
corresponding length scale of quantum physics for given mass scales
as shown in Table 2. Here `1 pc' denotes 1 parsec with 1 pc=$3\times 10^{16}$m.
\begin{center}
\renewcommand{\arraystretch}{1.7}
\begin{tabular}{|c |cc|}\hline
length & daily\hspace{8mm} & astronomy $\&$ \\
scale & life\hspace{8mm} & astrophysics \\ \hline
$1/M_{{\rm Planck}}$ & $10^{-35}$m & $10^{-52}$pc \\
$1/M_{{\rm EM}}$ & $10^{-19}$m & $10^{-36}$pc \\ \hline
\end{tabular}
\end{center}

Our spacetime signature is $(+,-,-,-,-)$ and definitions of the various
quantities we use are displayed in the following Table 3.
\begin{center}
\renewcommand{\arraystretch}{1.7}
\begin{tabular}{|c |c |}\hline
quantity & definition \\ \hline
Jacobian factor & $ g \equiv \det(g_{\mu\nu}) $ \\
connection           &   $\Gamma ^\mu _{\nu\rho} \equiv
\frac{1}{2}g^{\mu\sigma}(\partial_\nu g_{\sigma \rho}+\partial_\rho
g_{\sigma \nu} -\partial_\sigma g_{\nu \rho}) $          \\
covariant derivative of a contravariant vector & $ \nabla_\mu A^\nu
\equiv \partial_\mu A^\nu + \Gamma^\nu_{\mu\rho}A^\rho $      \\
Riemann curvature tensor & $
R^{\mu}_{\;\;\nu\rho\sigma} \equiv \partial_\rho \Gamma^{\mu}_{\sigma\nu}
-\partial_{\sigma} \Gamma^{\mu}_{\rho\nu}+
\Gamma^{\mu}_{\rho\tau}\Gamma^{\tau}_{\sigma\nu}
-\Gamma^{\mu}_{\sigma\tau} \Gamma^{\tau}_{\rho\nu} $\\
Ricci tensor  &  $ R_{\mu\nu} \equiv R^{\rho}_{ \; \mu\rho\nu}$ \\
curvature scalar  & $ R \equiv g^{\mu\nu}R_{\mu\nu}$ \\
Einstein tensor & $\displaystyle{G_{\mu\nu}\equiv R_{\mu\nu}
-\frac{g_{\mu\nu}}{2}R}$ \\
\hline
\end{tabular}
\end{center}

\setcounter{equation}{0}
\section{Einstein Equations and Geodesic Equations}

In this appendix we present detailed calculation of deriving
Einstein equations, geodesic equations, and Kretschmann invariant for the
metrics used in the description of Randall-Sundrum brane world scenarios
by using the formulas in Appendix A.

For the metric (\ref{cosmet}) of warp coordinates, nonvanishing
components of the connection are
\begin{eqnarray}\label{agabc} 
\Gamma^t_{tZ}=\Gamma^{x^{i}}_{x^{i}Z}=
A^{'},~~\Gamma^t_{x^{i}x^{i}}=e^{2b}\,\dot{b},
~~\Gamma^{x^{i}}_{tx^{i}}= \dot{b},~~
\Gamma^{Z}_{tt}=e^{2A}A^{'},~~\Gamma^{Z}_{x^{i}x^{i}}=-e^{2A+2b}A^{'}.
\end{eqnarray}
Nonvanishing components of the Riemann curvature tensor are
\begin{eqnarray}
&&R^{t}\,_{x^{i}tx^{i}}= -e^{2b}(e^{2A}A^{'2}-\dot{b}^{2}-\ddot{b}),~~
R^{t}\,_{ZtZ}=R^{x^{i}}\,_{Zx^{i}Z}=-A^{'2}-A^{''},\nonumber\\
&&R^{x^{i}}\,_{ttx^{i}}=-e^{2A}A^{'2}+\dot{b}^{2}+\ddot{b},~~
R^{Z}\,_{x^{i}x^{i}Z}=e^{2A+2b}(A^{'}+A^{''}),\nonumber\\
&&R^{x^{i}}\,_{x^{j}x^{i}x^{j}}= -e^{2b}(e^{2A}A^{'2}-\dot{b}^{2}),~~
R^{x^{i}}\,_{x^{j}x^{j}x^{i}}= e^{2b}(e^{2A}A^{'2}-\dot{b}^{2}),\nonumber\\
&&R^{Z}\,_{ttZ}=-e^{2A}(A^{'2}+A^{''}),
\label{arabcd}
\end{eqnarray}
and those of Ricci tensor are
\begin{eqnarray}\label{arab}
&&R_{tt}=(p+1)e^{2A}A^{'2}-p\dot{b}^{2}+e^{2A}A^{''}-p\ddot{b},
\nonumber\\
&&R_{x^{i}x^{i}}=-e^{2b}[(p+1)e^{2A}A^{'2}-p\dot{b}^{2}+e^{2A}A{''}
-\ddot{b}],~~
R_{ZZ}=-(p+1)(A^{'2}+A^{''}).
\end{eqnarray}
Finally the curvature scalar is
\begin{eqnarray}\label{ar}
R=2e^{-2A}\left[\frac{(p+1)(p+2)}{2}
e^{2A}A^{'2}-\frac{p(p+1)}{2}\dot{b}^{2}+(p+1)e^{2A}A^{''}
-p\ddot{b}\right].
\end{eqnarray}
From Eqs.~(\ref{arab})--(\ref{ar}),
nonvanishing components of the Einstein equations (\ref{ein}) are in
arbitrary $D$-dimensions
\begin{eqnarray}
G^{t}_{t} &=& \frac{p(p-1)}{2} \dot{b}^{2}e^{-2A}
-\frac{p(p+1)}{2} {A'}^{2}
-p {A''}
= -\frac{8\pi}{M_{*}^{p}} T^{t}_{\;t} + \Lambda \, ,
\label{acein1}\\
G^{i}_{i} &=& (p-1) \ddot{b}e^{-2A}
+\frac{p(p-1)}{2} \dot{b}^2 e^{-2A}
-\frac{p(p+1)}{2} {A'}^{2}
-p {A''}
= -\frac{8\pi}{M_{*}^{p}} T^{i}_{\;i} + \Lambda \, ,
\label{acein2}\\
G^{Z}_{Z} &=& p\ddot{b} e^{-2A} +
\frac{p(p+1)}{2} \dot{b}^2 e^{-2A}
-\frac{p(p+1)}{2} {A'}^2
= -\frac{8\pi}{M_{*}^{p}} T^{Z}_{\;Z} + \Lambda \, .
\label{acein3}
\end{eqnarray}
Simplifying the above equations (\ref{acein1})$\sim$(\ref{acein3}),
we have
\begin{eqnarray}
{(1-p)\ddot{b}} e^{-2A}&=&
-\frac{8\pi}{M_{*}^{p}}\left(T^{t}_{\;t} - T^{i}_{\;i}\right)
\label{seti}\, , \\
p(\ddot{b} + \dot{b}^{2})e^{-2A} + pA''
&=& -\frac{8\pi}{M_{*}^{p}} \left(T_{\;Z}^{Z} - T_{\;t}^{t} \right)
\label{sezt1}\, , \\
-\left(\frac{2}{p+1}\ddot{b}+\dot{b}^{2}\right)e^{-2A}+A^{'2}
&=&\frac{2}{p(p+1)}\left(\frac{8\pi}{M^{p}_{\ast}}T^{Z}_{\;Z}-\Lambda\right)
\label{sezi1}.
\end{eqnarray}
Once we turn off the time-dependence of the scale factor $b(t)$,
Eqs.~(\ref{seti})--(\ref{sezi1}) become Eqs.~(\ref{sezi})--(\ref{sezt}).

Structure of a fixed curved spacetime is usually probed by classical motions
of a test particle.
Once we obtain geometry of a brane world, then motions of a classical
test particle in the given background gravity $g_{AB}$ of the $D$-dimensional
bulk are described by geodesic equations
\begin{equation}\label{geo}
\frac{d^{2}x^{A}}{ds^{2}}
+\Gamma^{A}_{BC}\frac{dx^{B}}{ds}\frac{dx^{C}}{ds}=0,
\end{equation}
where the parameter $s$ is chosen by the proper time itself, a
force-free test particle moves on a geodesic.
For the metric with warp factor (\ref{cosmet}), nontrivial components
of the geodesic equations (\ref{geo}) are
\begin{eqnarray}
\frac{d^2 t}{ds ^2} + A^{'}\frac{dt}{ds}\frac{dZ}{ds}
+ \dot{b}e^{2b}\left(\frac{dx^{i}}{ds}\right)^2 = 0, \\
\frac{d^2 x^i}{ds^2} +A^{'}\frac{dx^i}{ds}\frac{dZ}{ds}
+ \dot{b} \frac{dt}{ds} \frac{dx^i}{ds}=0, \\
\frac{d^2 Z}{ds^2} + e^{2A}A^{'}\left(\frac{dt}{ds}\right)^2
-e^{2A+2b}A^{'}\left(\frac{dx^{i}}{ds}\right)^2 = 0.
\end{eqnarray}

The Kretschmann invariant is
\begin{equation}
R^{ABCD}R_{ABCD}=4p\!\left[(\ddot{b}+\dot{b}^{2})e^{-2A}\!\!\!
-A^{'2}\right]^{2}
\!\! +2p(p-1)\left(\dot{b}^{2}e^{-2A}\!\!\! -A^{'2}\right)^{2}
\!\!  +4(p+1)(A^{''}+A^{'2})^{2} ,
\end{equation}
which reduces to Eq.~(\ref{KrA}) in its static limit.

Let us repeat calculation for the Schwarzschild-type metric
\begin{equation}\label{metric1c}
ds^{2} = B(y) [dt^{2}-e^{2b(t)}dx^{i2}]-\frac{dy^2}{B(y)}.
\end{equation}
We have nonvanishing components of the connection
\begin{eqnarray}
&&\Gamma^t_{ty}=\Gamma^{x^{i}}_{x^{i}y}=\frac{B^{'}}{2B},~~
\Gamma^{y}_{yy}=-\frac{B^{'}}{2B},~~
\Gamma^t_{x^{i}x^{i}}=e^{2b}\dot{b},\nonumber\\
&&\Gamma^{x^{i}}_{tx^{i}}= \dot{b},~~
\Gamma^{y}_{tt}=\frac{1}{2}BB^{'},~~
\Gamma^{y}_{x^{i}x^{i}}=-\frac{e^{2b}}{2}BB^{'}.
\end{eqnarray}
Nonvanishing components of the Riemann curvature tensor are
\begin{eqnarray}
&&R^{t}_{x^{i}tx^{i}}=\frac{e^{2b}}{4}(4\dot{b}^{2}-B^{'2}+4\ddot{b}),~~
R^{t}\,_{yty}=R^{x^{i}}\,_{yx^{i}y}=-\frac{B^{''}}{2B}\nonumber\\
&&R^{x^{i}}\,_{ttx^{i}}=\frac{1}{4}(4\dot{b}-B^{'2}+4\ddot{b}),~~
R^{x^{i}}\,_{x^{j}x^{i}x^{j}}=\frac{e^{2b}}{4}(4\dot{b}^{2}-B^{'2}),\nonumber\\
&&R^{x^{i}}\,_{x^{j}x^{j}x^{i}}=-\frac{e^{2b}}{4}(4\dot{b}^{2}-B^{'2}),~~
R^{y}\,_{tty}=-\frac{1}{2}BB^{''},~~
R^{y}\,_{x^{i}x^{i}y}=\frac{e^{2b}}{2}BB^{''},
\end{eqnarray}
and those of the Ricci tensor are
\begin{eqnarray}
&&R_{tt}=\frac{1}{4}(-4p\dot{b}^{2}+pB^{'2}-4p\ddot{b}+2BB^{''}),~~
R_{x^{i}x^{i}}=\frac{e^{2b}}{4}(4p\dot{b}^{2}-pB^{'2}
+4\ddot{b}-2BB^{''}),\nonumber\\
&&R_{yy}=-\frac{(p+1)B^{''}}{2B}.
\end{eqnarray}
The curvature scalar is
\begin{eqnarray}
R=\frac{-4p(p+1)\dot{b}^{2}+p(p+1)B^{'2}-8p\ddot{b}+4(p+1)BB^{''}}{4B}.
\end{eqnarray}

Again, we read the $D$-dimensional Einstein equations (\ref{ein}) under this
metric
\begin{eqnarray}
G^{t}_{t} &=& \frac{p(p-1)}{2B}{\dot{b}}^2-\frac{p(p-1)}{8}
\frac{B'^2}{B} -\frac{p}{2}B''
= -\frac{8\pi}{M_{*}^{p}} T^{t}_{t} + \Lambda , \label{gtt}\\
G^{i}_{i} &=& \frac{p-1}{B}\ddot{b}+ \frac{p(p-1)}{2B}\dot{b}^{2}
-\frac{p(p-1)}{8}\frac{{B'}^2}{B} -\frac{p}{2}B''
= -\frac{8\pi}{M_{*}^{p}} T^{i}_{i} + \Lambda , \label{gii}\\
G^{y}_{y} &=& \frac{p}{B}\ddot{b}+\frac{p(p+1)}{2B}\dot{b}^{2}
-\frac{p(p+1)}{8}\frac{B'^{2}}{B}
= -\frac{8\pi}{M_{*}^{p}} T^{y}_{y} + \Lambda .  \label{gzz}
\end{eqnarray}
Then the simplified Einstein equations are
\begin{eqnarray}
\frac{(1-p)\ddot{b}}{B} &=&
-\frac{8\pi}{M_{*}^{p}}
\left(T^{t}_{t} - T^{i}_{i}\right) ,\label{etti} \\
\frac{4}{p}\left(\ddot{b} + \dot{b}^2 \right) + B^{'2} + 2B^{''}B &=&
-\frac{32\pi}{pM_{*}^{p}} B \left( T^{y}_{y} - T^{i}_{i} \right)
,\label{ettz1}\\
-4\left(\frac{2\ddot{b}}{p+1} + \dot{b}^2 \right) + B^{'^2}
&=& \frac{8B}{p(p+1)} \left( \frac{8\pi}{M_{*}^{p}} T^{y}_{y} -\Lambda
\right )
. \label{etzi1}
\end{eqnarray}
Their static limit coincides with Eqs.~(\ref{etzi})--(\ref{ettz}).

Nontrivial components
of the geodesic equations~(\ref{geo}) are given by
\begin{eqnarray}
\frac{d^2 t}{ds^2} + \frac{B^{'}}{2B}\frac{dt}{ds}\frac{dy}{ds}
+e^{2b}\dot{b}\left(\frac{dx^i}{ds}\right )^2 =0, \\
\frac{d^2 x^i}{ds} + \frac{B^{'}}{2B}\frac{dx^i}{ds}\frac{dy}{ds}
+ \dot{b} \frac{dt}{ds}\frac{dx^{i}}{ds} = 0, \\
\frac{d^2 y}{ds^2} + \frac{BB^{'}}{2} \left(\frac{dt}{ds}\right)^2
-e^{2b}BB^{'}\left(\frac{dx^i}{ds}\right)^2
-\frac{B^{'}}{2B}\left(\frac{dy}{ds}\right)^2 = 0.
\end{eqnarray}
Similarly, we read the Kretschmann invariant under the metric (\ref{metric1c})
\begin{equation}\label{bkret}
R^{ABCD}R_{ABCD}=\frac{1}{B^{2}}\left[
4p\left(\ddot{b}+\dot{b}^{2}-\frac{B^{'2}}{4}\right)^{2}
+2p(p-1)\left(\dot{b}^{2} -\frac{B^{' 2}}{4} \right)^{2}
+(p+1)B^{'' 2}
\right] .
\end{equation}
Its static limit reproduces Eq.~(\ref{KrB}).

\setcounter{equation}{0}
\section{Small Gravitational Fluctuations}
In this appendix, we will give the detailed derivation for
Eq.~(\ref{fili}). Let us consider the variation of the Einstein
equations (\ref{ein})
\begin{eqnarray}\label{lieq2}
&&\hspace{-15mm}\delta\left(
G_{AB}=-\frac{8\pi}{M^{p}_{\ast}}T_{AB}
+\Lambda g_{AB} \right) ,
\end{eqnarray}
where
\begin{equation}\label{dric}
\delta G_{AB}\equiv
\delta(R_{AB} - \frac{1}{2} g_{AB}R )
= \delta R_{AB} - \frac{1}{2} \delta g_{AB} R.
\end{equation}
The variation of Ricci tensor in Eq.~(\ref{dric}) is given by
\begin{equation}\label{der1}
\delta R_{AB} = \nabla_B \delta C^C_{CA} -\nabla_C \delta C^C_{AB}\, ,
\end{equation}
where, for small fluctuations, $C^{A}_{BC}$ is
\begin{equation}\label{der2}
\delta C^D_{AB}\equiv \frac{1}{2} g^{AD}( \nabla_B \delta g_{CD} +
\nabla_C \delta g_{BD} - \nabla_D \delta g_{BC} ).
\end{equation}

From here on we derive Eqs.~(\ref{der1})--(\ref{der2}). If we consider
variation of the covariant derivative for a vector as
\begin{equation}
(\delta \nabla_A )V_B = \tilde{\nabla}_A V_B - \nabla_A V_B
\equiv - C^C_{AB}V_C \, ,
\end{equation}
where the tilde over the covariant derivative denotes
the quantity calculated on the basis of the
perturbed metric $\tilde{g}_{AB} = g_{AB} + \delta g_{AB}$.
After some straightforward calculations of $C^A_{BC}$ from its
definition, we have
\begin{equation}
- C^C_{AB}V_C =(-\tilde{\Gamma}^C_{AB} + \Gamma^C_{AB})V_C,
\end{equation}
so that
\begin{equation}\label{cabc}
C^C_{AB}= \frac{1}{2} \tilde{g}^{CD}( \nabla_A
\tilde{g}_{BD} + \nabla_B \tilde{g}_{AD} - \nabla_D
\tilde{g}_{AB} ).
\end{equation}
For small gravitational fluctuations, Eq.~(\ref{cabc}) coincides with
Eq.~(\ref{der2}).

From the definition of Riemann tensor $[\nabla_A ,
\nabla_B] V_C \equiv V_D R^D_{CBA}$, we obtain an expression of small
variation of the Riemann curvature tensor
\begin{equation}
V_D \delta R^D_{CBA} = \delta([\nabla_A , \nabla_B]) V_C \\
=(\nabla_B \delta C^D_{AC} -\nabla_A \delta C^D_{BC})V_D \, ,
\end{equation}
which leads to
\begin{equation}
\delta R^D_{CBA} = \nabla_B \delta C^D_{AC} -\nabla_A \delta
C^D_{BC} \, .
\end{equation}
Contraction of two indices provides that of the Ricci tensor in
Eq.~(\ref{der1}).

Now specific computation of fluctuation equations for
Eqs.~(\ref{fmet})--(\ref{rsga}) is in order.
Variation of the Ricci tensor (\ref{der1}) is calculated by using the
expression of $C^P_{MN}$ (\ref{der2})
\begin{eqnarray}
\delta R_{AB} &=& \frac{1}{2}\nabla_B \nabla_A \delta g^C_C -
\frac{1}{2} \nabla^C( \nabla_B \delta g_{AC} + \nabla_A \delta
g_{BC} -\nabla_C \delta g_{AB})\, , \\
\delta R_{\mu\nu} &=&\frac{1}{2}\nabla^2 h_{\mu\nu}+ \frac{1}{2}
\nabla_{\mu}\nabla_{\nu} h - \frac{1}{2} \nabla^A(\nabla_{\mu}
h_{\nu A}+\nabla_{\nu} h_{\mu A} ) \, ,
\end{eqnarray}
where $h_{\mu\nu}=\delta g_{\mu\nu}$ and $h=h^{\mu}_{\mu} =
\delta g^{\mu}_{\mu}$. Note that
\begin{equation}
U'=\frac{dU}{dZ}, \qquad U''
=\frac{d^{2}U}{dZ^{2}}.
\end{equation}
Under the transverse-traceless gauge (\ref{rsga}), we obtain an expression
for variation of the Ricci tensor
\begin{equation}
\delta R_{\mu\nu} = \frac{1}{2} \left( g^{\rho\sigma}
\partial_{\rho}\partial_{\sigma} - \partial^2_Z \right) h_{\mu\nu} -
\frac{1}{2} \left( \frac{U'}{U}\right)^2 h_{\mu\nu}.
\end{equation}
Here we used nonvanishing components of the connection before turning
on the fluctuations, which should have only one $Z$ index
\begin{eqnarray}
\Gamma^{Z}_{\mu\nu}=\frac{1}{2}U'\eta_{\mu\nu},\qquad
\Gamma^{\mu}_{Z\nu}=\frac{1}{2}\left(\frac{U'}{U}\right)\delta^{\mu}_{\;\nu}.
\end{eqnarray}
Substituting the scalar curvature
\begin{equation}
R= -4 \left(\frac{U''}{U}\right) -
\left(\frac{U'}{U}\right)^2 ,
\end{equation}
we obtain the variation of gravity
part, the left-hand side of the Einstein equations (\ref{dric})
\begin{eqnarray}
\delta(R_{\mu\nu} - \frac{1}{2} g_{\mu\nu} R )
&=& \delta R_{\mu\nu} - \frac{1}{2} h_{\mu\nu} R \\
&=&\frac{1}{2} \left( g^{\rho \sigma}
\partial_{\rho}\partial_{\sigma} - \partial^2_Z \right) h_{\mu\nu} +
2 \left( \frac{U''}{U}\right)^2 h_{\mu\nu}.
\label{left}
\end{eqnarray}
Variation of the matter part, the right-hand side of the Einstein
equations, is
\begin{equation}
\delta \left(-\frac{8\pi}{M^{p}_{\ast}}T_{AB} +\Lambda g_{AB}
\right) = -\frac{8\pi}{M^{p}_{\ast}}\delta T_{AB} +\Lambda \delta
g_{AB}.
\label{right}
\end{equation}
From form of the matter source (\ref{em20}), we read
\begin{equation}
\delta T_{ZZ} =0, \qquad
\delta T_{\mu\nu} =\frac{M^{p}_{\ast}}{8\pi} 2pk \delta (Z) h_{\mu\nu},
\end{equation}
where we used the relation $T_{AB} = g_{AC} T^C_B$.
By comparing Eq.~(\ref{left}) and Eq.~(\ref{right}), we finally arrive at
the Einstein equations for the small gravitational
fluctuations (\ref{fili}).

\select{}{\end{multicols}}

\begin{thebibliography}{100}
\bibitem{ADD} N. Arkani-Hamed, S. Dimopoulos and G. Dvali, Phys. Lett. B {\bf
429}, 263 (1998) [hep-ph/9803315]; Phys. Rev. D {\bf 59}, 086004
(1999), {\tt[hep-ph/9807344]}; I. Antoniadis, N. Arkani-Hamed, S. Dimopoulos
and G. Dvali, Phys. Lett. B {\bf 436}, 257 (1998), {\tt[hep-ph/9804398]}.
\bibitem{RS1} L. Randall and R. Sundrum, Phys. Rev. Lett. {\bf 83}, 3370
(1999), {\tt [hep-ph/9905221]}.
\bibitem{RS2} L. Randall and R. Sundrum, Phys. Rev. Lett. {\bf 83}, 4690
(1999), {\tt [hep-th/9906064]}.
\bibitem{Rev} For a review, see
V.A. Rubakov, Phys. Usp. {\bf 44}, 871 (2001), {\tt[hep-th/0104152]};
R. Dick, Class. Quant. Grav. {\bf 18}, R1 (2001), {\tt[hep-th/0105320]};
D. Langlois, Prog. Theor. Phys. Suppl. {\bf 148}, 181 (2003),
{\tt[hep-th/0209261]}.
\bibitem{KK} N. Kaloper, Phys. Rev. D {\bf 60}, 123506
(1999), {\tt [hep-th/9905210]}; T. Nihei, Phys. Lett. B {\bf 465}, 81
(1999), {\tt [hep-ph/9905487]};
H.B. Kim and H.D. Kim, Phys. Rev. D {\bf 61}, 064003
(2000), {\tt [hep-th/9909053]}.
\bibitem{GW} W.D. Goldberger and M.B. Wise, Phys. Rev. D {\bf 60}, 107505
(1999), {\tt[hep-ph/9907218]}; Phys. Rev. Lett. {\bf 83}, 4922 (1999),
{\tt[hep-ph/9907447]}.
\bibitem{GRS} R. Gregory, V.A. Rubakov, and S.M. Sibiryakov,
Phys. Rev. Lett. {\bf 84}, 5928 (2000), {\tt[hep-th/0002072]}.
\bibitem{BDL} P. Binetruy, C. Deffayet, and D. Langlois,
Nucl. Phys. B {\bf 565}, 269 (2000), {\tt[hep-th/9905012]};
J.~M.Cline and C. Grojean, Phys. Rev. Lett. {\bf 83}, 4245 (1999),
{\tt[hep-th/9906523]};
P. Binetruy, C. Deffayet, U. Ellwanger, and D. Langlois,
Phys. Lett. B {\bf 477}, 285 (2000), {\tt[hep-th/9910219]}.
\bibitem{Od} O. DeWolfe, D.Z. Freedman, S.S. Gubser, and A. Karch, Phys.
Rev. D {\bf 62}, 046008 (2000), {\tt [hep-th/9909134]}.
\bibitem{ABN} R. Altendorfer, J. Bagger, and D. Nemeschansky,
Phys. Rev. D {\bf 63} 125025 (2001), {\tt[hep-th/0003117]};
A. Falkowski, Z. Lalak, and S. Pokorski, Phys. Lett. B {\bf 491}, 172 (2000),
{\tt[hep-th/0004093]}.
\bibitem{CCV} C.S. Chan, P.L. Paul, and H. Verlinde,
Nucl. Phys. B {\bf 581}, 156 (2000), {\tt[hep-th/0003236]}.
\bibitem{CK} A.G. Cohen and D.B. Kaplan, Phys. Lett. B {\bf 470}
52 (1999), {\tt [hep-th/9910132]}; R. Gregory, Phys. Rev. Lett. {\bf 84} 2564
(2000) {\tt[hep-th/9911015]}.
\bibitem{APR} N. Arkani-Hamed, M. Porrati, and L. Randall,
JHEP {\bf 0108}, 017 (2001), {\tt[hep-th/9910132]}.
\bibitem{ADKS} N. Arkani-Hamed, S. Dimopoulos, N. Kaloper, and
R. Sundrum, Phys. Lett. B {\bf 480}, 193 (2000), {\tt[hep-th/0001197]};
S. Kachru, M.B. Schulz, and E. Silverstein, Phys. Rev. D {\bf 62}, 045021
(2000), {\tt[hep-th/0001206]}.
\bibitem{KIS} H. Kodama, A. Ishibashi, and O. Seto, Phys. Rev. D {\bf 62},
064022 (2000), {\tt[hep-th/0004160]}.
\end{thebibliography}
\end{document}